# Demand-side policies for power generation in response to the energy crisis: a model analysis for Italy

Alice Di Bella[1*], Massimo Tavoni[1,2]

[1]*RFF-CMCC European Institute on Economics and the Environment (EIEE), Centro Euro-Mediterraneo sui Cambiamenti Climatici, Italy*
[2]*Politecnico di Milan, Italy*

[*]E-mail: alicedb7@gmail.com

**Abstract**

In order to mitigate the impacts of the energy crise, the European Union has proposed various measures. For the power sector a directive prescribes a shift of 5% of the demand in 10% of the peak hours, plus a voluntary 10% overall demand reduction. Here we use a power system model to quantify the implications of this policy for the Italian power sector, as it stands today and under the transformation required to meet the climate goals of the Fit-for-55. We find that policymakers would need to incentivize electricity consumption in the middle of the day while discouraging it in the early morning and late afternoon. We also highlight the benefits of the decarbonization strategy in the context of uncertain gas prices: for a gas price at or above 50€/MWh, power generation through gas is reduced by more than one third, approaching what needed to comply with the Fit-for-55. Finally, we quantify the value of demand side management strategies to curb fossil resource consumption and to reduce curtailed electricity under a high renewable scenario.

1. Introduction

Due to recent political development, global gas prices, particularly European ones, have suffered consistent and unexpected volatility. Natural gas in Europe is usually traded at the virtual trading point in the Netherlands (TTF, Title Transfer Facility) and, for the last five years, the average gas cost in this market oscillated between 10 and 30 €/MWh of thermal energy[1]. Prices started increasing at the end of 2021, following a moderate economic rebound after the COVID-19 pandemic, surpassing the value of 50€/MWh. Still, with the beginning and the evolution of the Russian invasion of Ukraine, prices went up uncontrollably: on February 24th, they broke the 100€/MWh wall and at the end of August they reached the record of 340 €/MWh. In Europe, this crisis is shaking the basis of the current energy system, which now appears vulnerable and dangerously dependent on imports from the Russian Federation.

Russia is one of the biggest natural gas exporters in the world: in 2021, according to the Bank of Russia[2], the value of national exports of this resource was 7.32 billion dollars for LNG and 5.5 billion for gas through pipelines. European Union is particularly dependent on Russian gas due to past policies that led to a scarce differentiation of sources. In 2021, the EU imported 45%[3] of the natural gas consumed in the Union (155 bcm) from the Russian Federation; Italian energy security is also at risk, since 38%[4] (29 bcm) of our natural gas comes from Russia. In 2022 the origin of European gas import necessarily changed, enhancing trade of LNG with the US, Qatar and Nigeria, but Russia still held a share of around 30% of the market in the first half of 2022[5].

This price crisis affects many European families and businesses, thus various EU country leaders are taking measures to protect their citizens from extra expenses. Recently, the European Council[6] has proposed a measure to increase energy security in the power sector: they agreed to a voluntary overall reduction target of 10% of gross electricity consumption and a mandatory reduction target of 5% of the electricity consumption in peak hours. Member states will identify 10% of their peak hours between December 1st 2022 and March 31st 2023, during which they will reduce the demand. Member states will be free to choose the appropriate measures to reduce consumption for both targets in this period. Substantially, this proposal consists of two different operations: first, a shift in electricity consumption from the peak hours (i.e. the hours with a higher load, when it can be more costly to satisfy the demand), which basically can be represented as a load shift; the other one is directly a demand reduction for four months of the year.

This paper the impact of these two measures on gas consumption in Italy, thanks to a power model of the whole Italian electricity system developed with the open-source modelling framework Oemof (see next section for details). Chapter 2 will describe the methodology for developing the model and retrieving. Considering recent gas price trends, it is also interesting to investigate synergies with the energy transition. These high prices could make the shift to a cleaner way to produce electricity a cost-efficient intervention to reduce dependency on gas imports, while simultaneously achieving legally binding objectives of the Fit-for-55 package from the EU Commission. As the Commission President, Ursula von der Leyen, affirmed during COP27 in Sharm el-Sheikh, "*The global fossil fuel crisis must be a game changer*"[7]. Finally, the potential of load shifting and demand side management in the energy transition will be studied, since it could lead to the same goals at lower costs, deploying a case study for Italy in 2030. The results of these three research processes will be outlined and discussed in the third chapter. Finally, the last chapter will present some recommendations for policymakers.

**2. Materials and methods**

A power system model of the Italian electricity system was employed to answer the three research questions presented above, developed with Oemof (*Open Energy MOdelling Framework*)[8], an open-source energy modelling tool in Python. This framework enables the creation of an energy network through the library *Oemof-network* and then solves the energy balances with *Oemof-solph*. The solver, in this case, was Gurobi, but Oemof can use other open-source solvers. The Italian power system is described with an hourly resolution, divided into the seven market zones defined by the transmission system operator Terna[9,10]. Oemof allows the modeler to design different elements of an electric grid and here they will be listed briefly; for a more thorough description, please refer to Di Bella et al.[11].

- Energy demand: peak value and a normalized time series per each region
- Variable RES plants: availability distributions, installed capacity and maximum potential
- Power plants and storages: installed capacity, efficiency, fuel used and eventually a maximum potential
- Transmission lines: capacity linking different nodes
- Technologies investment and operation costs
- Emission factors for fossil fuel sources

For this study, in the Oemof model of the Italian power sector, two different optimization methods are utilized based on the research topic investigated: dispatch and investment optimization. The first one is employed to analyze the practical impact of the Council directive, which should be applied starting in December at the current electric system, so no relevant extra investment in capacity is envisioned. The second one is applied for the two other questions.

The crucial factor in this study is Demand Side Management, which is implemented in Oemof thanks to the DSM component[12], which offers three different possible approaches: a standard one (called simply 'oemof'), one based on Gils et al.[13] ('DLR') and one based on Zerrahn and Schill[14] ('DIW'). For this analysis, the standard method is enough to investigate all the aspects of the Council proposal. The major constraint of a dispatch optimization is to meet the electric demand $D_{t,n}$ at any time step t for each node n, with electricity generation from each source s in the node n $E_{t,n,s}$, considering also the charge $E_{t,n,st}^{charge}$ and discharge energy $E_{t,n,st}^{discharge}$ of the storage technology st in node n and the transmission losses $E_{t,p}^{transmission\ loss}$ for every powerline p (sometimes with excess of generation $E_{t,n}^{excess}$).
DSM changes the demand side of the equation: demand is multiplied for the max load considered in the node $Dmax_n$, increased by a possible extra load from $DSM_{t,n}^{up}$ or reduced by $DSM_{t,n}^{down}$, as in Eq. 1.

$$\sum_{st}^{St}\sum_{p}^{P}\sum_{s}^{S}\left(E_{t,n,s} + E_{t,n,st}^{discharge} - E_{t,n,st}^{charge} - E_{t,p}^{transmission\ loss}\right) = D_{t,n} \cdot Dmax_n + DSM_{t,n}^{up} - DSM_{t,n}^{down} \quad (1)$$

$DSM_{t,n}^{up}$ and $DSM_{t,n}^{down}$ are subject to limitations since their values need to be lower than a fraction ($\%D_{t,n}^{up\ max}$ or $\%D_{t,n}^{down\ max}$) of the actual demand $D_{t,n}$ per each node n and at each time step t. (see Eq. 2 and 3)

$$DSM_{t,n}^{up} \leq D_{t,n} \cdot \%D_{t,n}^{up\ max} \quad (2)$$

$$DSM_{t,n}^{down} \leq D_{t,n} \cdot \%D_{t,n}^{down\ max} \quad (3)$$

The shifts up and down need to be compensated within a certain time interval $\tau$ after the moment when the first shift happens $t_{shift}$ (Eq. 4). In this study, the shift interval is assumed to be 24 hours, imagining to compensate a daily load.

$$\sum_{t=t_{shift}}^{t_{shift}+\tau} DSM_{t,n}^{up} = \sum_{t=t_{shift}}^{t_{shift}+\tau} DSM_{t,n}^{down} \quad (4)$$

This feature could have extra operation costs, $CostDSM_{t,n}^{up}$ and $CostDSM_{t,n}^{down}$, that need to be added to the usual costs of generation $vc_{n,s}$ to produce electricity at each time step t by the source s $E_{t,n,s}$ (see Eq. 5). For this paper the operation costs are set to null to see the optimization of DSM without any cost constraint.

$$C^{operation} = \sum_{t=1}^{T}\sum_{n=1}^{N}\sum_{s=1}^{S}(E_{t,n,s} \cdot vc_{n,s} + DSM_{t,n}^{up} \cdot CostDSM_{t,n}^{up} + DSM_{t,n}^{down} \cdot CostDSM_{t,n}^{down}) \quad (5)$$

There are also various other constraints, like considering the maximum power for each generator unit and the storage balance. Here they will be omitted to stay brief, but a detailed explanation can be found in the paper from Prina et al.[15].

For the second part, evaluating the optimal power sector for different gas prices, the model of the Italian system will perform an investment optimization, thus also including the price for installing new capacity of different technologies in the cost equation. The total system annual costs $C^{total}$, which must be minimized, are calculated as in Eq. 6.

$$Min\ C^{total} = Min\ C^{operation} + \sum_{n=1}^{N}(\sum_{s=1}^{S} C_s \cdot P_{n,s}^{added} + \sum_{st=1}^{ST} C_{st} \cdot E_{n,st}^{added} + \sum_{p=1}^{P} C_p \cdot P_p^{added}) \quad (6)$$

$C_s$ = resource s expansion capital cost
$C_{st}$ = storage tech. st expansion capital cost
$C_p$ = powerlines expansion capital cost

$P_{n,s}^{added}$ = capacity added for source s in node n
$E_{n,st}^{added}$ = capacity added for storage st in node n
$P_p^{added}$ = capacity added for powerline p

In the last case study, the role of DSM in decarbonization pathways is evaluated, so a constraint for carbon dioxide emissions is employed to set the model to perform the needed investment to decarbonize the power system in Italy. The Oemof model is designed to run Single-Objective evaluations, minimizing the system's total costs for the chosen timeframe, but it also offers the possibility to add extra constraints for the power network. In this case, the optimization achieves the minimum total cost $C^{total}$ while limiting the total amount of carbon dioxide released $CO_2^{total}$ with a chosen value epsilon $\epsilon_i$ which represents the limit for absolute emissions of the system (see Eq. 7).

$$Min\ C^{total} \quad while\ CO_2^{total} \leq \epsilon_i \quad (7)$$

The total value of $CO_2$ emissions is calculated as reported in Eq. 8.

$$CO_2^{total} = \sum_{t=1}^{T}\sum_{n=1}^{N}\sum_{s=1}^{S} P_{t,n,s}^{fossil\ source} \cdot co_2^{factor}{}_s \leq \epsilon_i \quad (8)$$

t = analyzed time step, from 1 to T
n = node considered, from 1 to N
s = generation source, from 1 to S
$P_{t,n,s}^{fossil\ source}$ = thermal power supplied by the commodity source s in node n at time step t

$co_2^{factor}{}_s$ = emission factor for technology s [Mton/$CO_2$]
$\epsilon_i$ = limit for the total amount of $CO_2$ emitted in total by the system

**Italian power system**

The Italian electric system in 2021 is the study's starting point, so the model is validated on the 2021 electricity generation[16]. As already said, the optimization is performed with an hourly resolution for 7 Italian market zones.
In Figure 1, a scheme for the operation of the model is represented: in Oemof, renewable resources can directly produce electricity, while commodities like fossil fuels, reservoir hydro and import pass through a transformer, with the corresponding efficiency. The net electricity imported from abroad, reduced by the national transmission losses, is considered in the import component. It is assumed to be a generation source only in the North region since France is by far the primary importer. Pumped Hydro Storage (PHS) is the only storage source for the Italian power system in 2021.

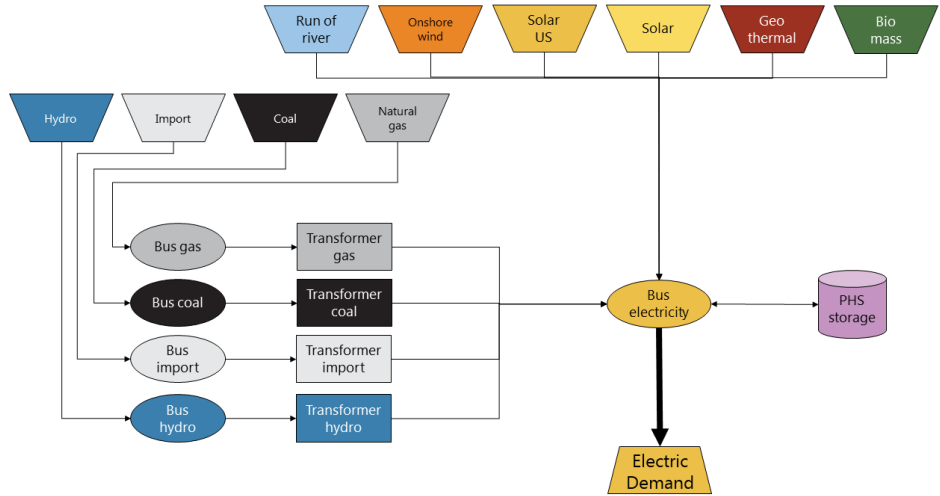

*Figure 1 – Oemof operational scheme for the Italian power sector in 2021*

Table 1 illustrates existing capacities for technologies in 2021, while in Table 2, there are the costs assumptions and in Table 3, capacities for transmission lines as of today and according to Terna 2021 Development Plan[17]. Input data for electric demand[16], efficiencies[15,18,19], transmission lines costs[20] and maximum potential for RES[21] are obtained from different data sources.

| [GW] | Coal[22] | Gas[23] | Rooftop PV[24] | Utility Scale PV[24] | On-wind[24] | Bio energy[24] | ROR[25] | Hydro[25] | PHS [GWh][25] |
|---|---|---|---|---|---|---|---|---|---|
| **North** | 1.3 | 27.7 | 9.4 | 1.5 | 0.2 | 2.3 | 4.4 | 2.2 | 340.0 |
| **Centre-North** | 0 | 2.6 | 1.9 | 0.3 | 0.2 | 0.4 | 0.2 | 0 | 0 |
| **Centre-South** | 2.0 | 7.5 | 2.7 | 1.3 | 2.2 | 0.5 | 1.1 | 0.1 | 176.5 |
| **South** | 2.6 | 4.9 | 2.7 | 0.9 | 4.8 | 0.5 | 0 | 0 | 0 |
| **Sardinia** | 1.1 | 1.1 | 0.5 | 0.6 | 1.1 | 0.1 | 0 | 0 | 12.5 |
| **Sicily** | 0 | 5.4 | 1.1 | 0.6 | 2.0 | 0.1 | 0 | 0 | 31.0 |
| **Calabria** | 0 | 3.6 | 0.5 | 0.1 | 1.2 | 0.2 | 0.1 | 0.3 | 0 |

*Table 1 – Existing capacity of generation and storage technology per market zone in Italy, 2021*

| | CAPEX [2021€/kW or 2021€/kWh] | FOM [%capex/y][19] | OPEX [2021€/MWh][19] | Lifetime [y][19] |
|---|---|---|---|---|
| **Utility scale PV** | 431[26] | 3 | 0.01 | 25 |
| **Rooftop PV** | 726[27] | 2 | 0.01 | 25 |
| **Onshore wind** | 1165[28] | 2.45 | 2.3 | 25 |
| **Offshore wind** | 2417[28] | 2.3 | 2.7 | 30 |
| **Li-ion batteries** | 239[29] (system costs[30]) | 0 | 0 | 15 |
| **Electrolyzers** | 748[31] | 2 | 0 | 20 |
| **Fuel cells** | 456[32] | 3 | 0 | 30 |
| **H2 storage** | 22.6[33] | 0 | 0 | 20 |

*Table 2 – Costs assumptions for Italy 2021*

| Transmission lines [1 – 2] | 2021 capacity towards 1 | 2021 capacity towards 2 | 2030 capacity towards 1 | 2030 capacity towards 2 |
|---|---|---|---|---|
| **North – Centre North** | 2700 | 3900 | 3100 | 4300 |
| **Centre North – Centre South** | 2400 | 2500 | 2800 | 2900 |
| **Centre South – South** | 4600 | 2000 | 5550 | 2000 |
| **Centre South – Sardinia** | 900 | 720 | 900 | 720 |
| **Centre North – Sardinia** | 395 | 315 | 1095 | 1015 |
| **Sicily - Sardinia** | 0 | 0 | 800 | 800 |
| **Centre North – Sicily** | 0 | 0 | 700 | 700 |
| **South - Calabria** | 2350 | 1100 | 2350 | 1100 |
| **Sicily - Calabria** | 1100 | 1200 | 1750 | 1200 |

*Table 3 – Capacity of transmission lines in 2021 and assumed in 2030*

Eight decision variables for each market zone are taken into account for the expansion capacity optimization: for renewable generation, onshore and offshore wind, rooftop and utility scale photovoltaics capacities, for storage, li-ion batteries, electrolyzers, fuel cells and hydrogen tanks. Transmission line capacities can also be expanded in both directions, for a total of 74 decision variables.

## 3. Results

1) **What is the impact of the European proposal of cutting electricity demand and shifting it during peak hours on gas consumption, system costs and emissions?**

For the evaluation of the effects of the new measures suggested by the European Council on the Italian power sector, the Oemof component for Demand Side Management is implemented. The proposal is supposed to be applied to the electrical demand between December 2022 and March 2023, so the study will examine that period with an hourly time granularity, using the same load for the period December 2021 - March 2022. The cost of DSM is set to zero: firstly, it is complicated to assign a price to this mechanism and secondly, the aim is observing the consequences of a policy already affirmed and that should be put in place, regardless of the costs. The Council proposal refers to a 5% reduction in demand during load peaks and a voluntary 10% demand reduction for four months. The effects of these two measures will be analyzed separately and then combined.

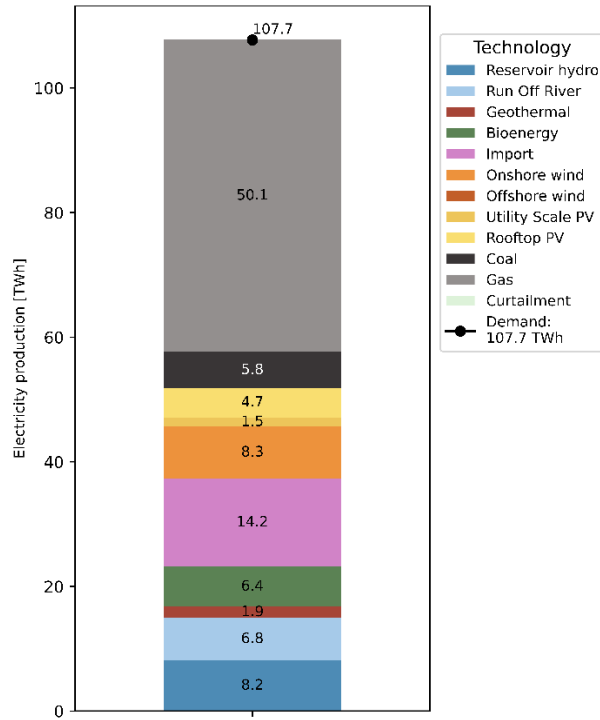

*Figure 2 – Italian electricity mix December to March*

First of all, the energy mix during these months is replicated in the model, as a reference scenario to compare the effects of DSM. A dispatch run of the Italian power system is performed, taking existing capacities and the load demand for the same months one year before. The results of the energy mix are displayed in Figure 2, where it is evident how gas plays a huge role in the Italian power sector.

After that, to reflect the first part of the EC measure, the DSM parameters for the percentage of demand that can be reduced or increased at each time step ($\%D_{t,n}^{up\ max}$ or $\%D_{t,n}^{down\ max}$) are set to 5% of the total original demand. The results of this optimization are not remarkably diverse from the current electricity mix, but the main differences are shown in Table 4: 9.6 GWh reduction in electricity from gas (corresponding to 17.4 GWh of resource gas), 3.6 kton less of $CO_2$ produced and 5.3 million euros saved in 4 months. These numbers are negligible compared to the absolute values of the corresponding parameters, since annual emissions from the power sector were 81.1 Mton of $CO_2$ in 2019[34], and it is also not particularly crucial for energy security, since natural gas imports from Russia consist of around 307 TWh of energy[4].

This part of the Council measure probably has a different goal: shifting load from peak hours can be a way to reduce volatility in gas prices and stop rises of this resource cost caused by demand-supply mechanisms. When the load demand peaks and renewable capacity is not enough to satisfy it, the system must dramatically meet that load with a flexible resource like gas. In those moments, prices can increase significantly, driven by a pressing request. Shifting the load during peak hours reduces the urgency to produce electricity with natural gas, thus it could be a tool to control prices instead of a way to decarbonize and reduce dependence on Russia. It is also important to highlight that DSM acts as a flexibility measure in the system, thus decreasing the need to store energy in PHS. In the current context, these results are not particularly relevant, but it could be interesting to study the role of this approach in a power system with a high share of RES. In the third part of this chapter[14], the impacts of DSM on decarbonization scenarios in 2030 will be outlined. The effects of DSM on today's system are not extraordinary mainly for two reasons: first, following the European Council directive, demand available for DSM is just a minor part of the total. Second, the current power system does not have enough RES to generate a relevant overgeneration, which can create the opportunity to shift a load in an hour when there is excess electricity production.

|  | Gas [GWh] | PHS [GWh] | Annual system cost [M€] | Emissions [tonCO2] |
|---|---|---|---|---|
| Today | 50136.9 | 8.1 | 27977.7 | 24327.5 |
| With 5% load shift | 50127.3 | 1.2 | 27972.4 | 24323.9 |
| With 10% demand red | 39391.6 | 48.2 | 22049.2 | 20387.2 |
| With 10% demand red and 5% load shift | 39375.4 | 19.6 | 22040.3 | 20382.2 |
| Diff % 10% dem red wrt today | -21.4% | +494.4% | -21.2% | -16.2% |
| Diff % 10% dem red and 5% load shift wrt today | -21.5% | +142.0% | -21.2% | -16.2% |

*Table 4 – Comparison of today's electricity mix, optimization of the system with 5% load shift, optimization of the system with 10% demand reduction and optimization with 10% reduction and 5% load shift at the same time (and difference with respect to today's system).*

The second part of the Council directive consists of a voluntary reduction of electricity consumption of 10% of the demand. Obviously, not consuming means reducing imports of natural gas (which has the highest operation costs), fewer greenhouse gases emitted and less money spent. This proposal is not so straightforward to implement for decision-makers since it will mean limiting the residential and industrial consumption of citizens and firms. Putting it into practice will require a complete analysis of the most influential and socially accepted ways to impose and enforce this policy. This case study is modeled by evenly reducing the load by 10% for every hour. In Table 4 it is visible, though, that the outcomes are quite interesting: a 10% demand reduction does not lead to a linear 10% reduction in gas consumption and a 10% decrease in emissions of $CO_2$, but it shows a nonlinear trend, boosting more the cutback.

This effect is possible thanks to the enhanced utilization of PHS, which stores almost five times more electricity than in the reference case. When the reduced demand in some hours is lower than the RES production, energy is stored as water in the upper basin. In peak hours, that water goes down the PHS turbine and produces electricity instead of turning on gas power plants. This technology has been underutilized recently[35] mainly due to scarce economic convenience for the plants owners, but it has been used way more in the past. PHS is still an appealing opportunity for flexibility in the Italian electric system for the next future, while more detailed analyses will be required to consider it for long-term scenarios. In fact, climate change could lead to drought and lower water availability, which could be directed more to satisfy drinking needs and agriculture, leaving less space for electricity generation.

Finally, it is possible to combine both measures (5% peak load shift and 10% demand reduction) and thus consider the impact of the complete implementation of the EC directive. Figure 3 compares the reference scenario for the current system and the three case studies. It confirms that the only effect of DSM applied to the existing power system is to reduce the load on PHS (Table 4). In contrast, to effectively reduce gas consumption, demand reduction appears to be a key factor.

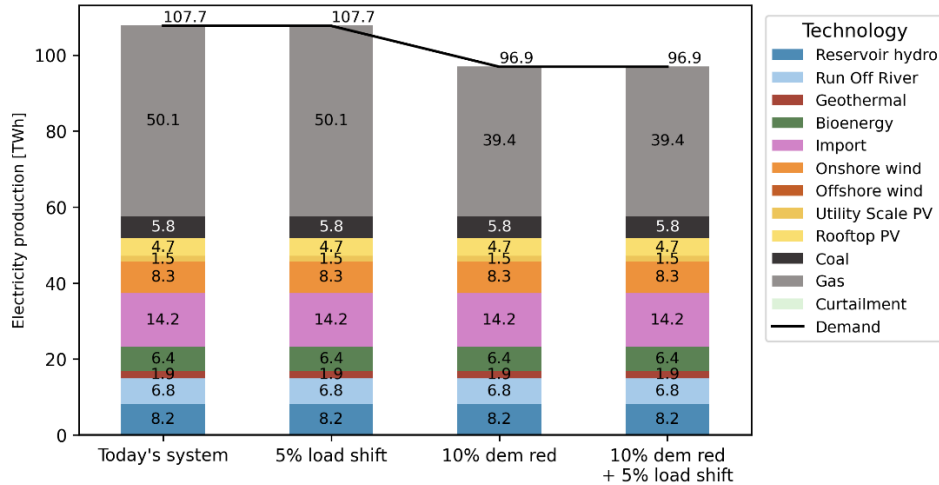

*Figure 3 – Comparison of the electricity mix in the current system and three case studies (5% DSM, 10% demand reduction, both)*

Member states also have the task to identify the 10% of their peak hours in the four months when the electricity request will be reduced by 5%. The model optimizes the hours when it is more convenient to reduce the demand and hours when it makes sense instead to increase the power demand; thus, it offers riveting insights to select those hours. For each market zone, the 10% largest load shifts up and down are taken, aggregated for the hour of the day when they happen and then aggregated again to have a national result, which is more useful for policy purposes. The total number of shifts per hour of the day, taking the largest 10%, is displayed in Figure 4, positive if it is more convenient to increase the load and negative for the opposite.

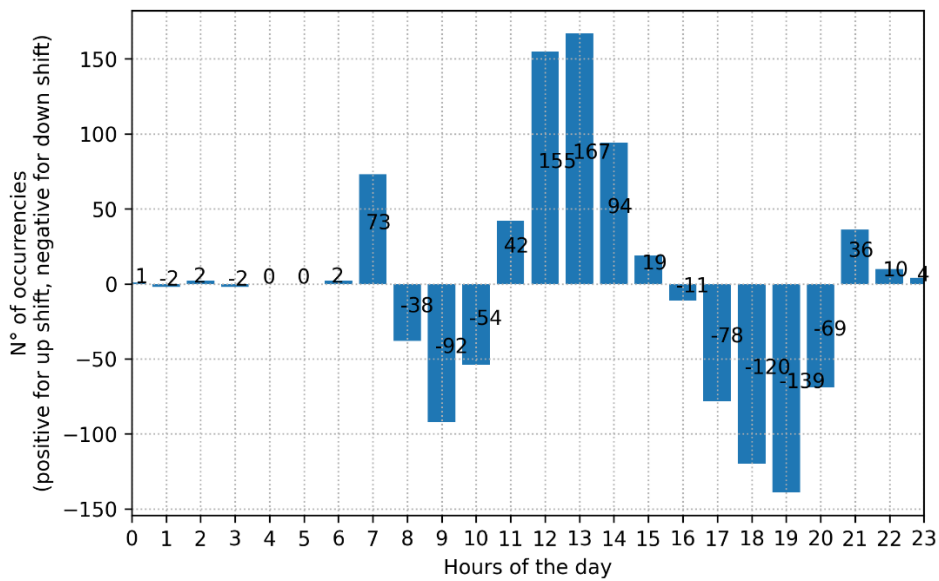

*Figure 4 – Number of shifts up (positive) or down (negative) in 10% of the total hours at national level*

The model pushes the load when there is excess of solar electricity, so during the middle of the day (between 11 and 14), while decreasing it before (8 to 10) and later (17 to 20), when there is less sun but more electricity request. A spark in upshift is also present at 7 and 21; this behavior could be led by the fact that these are moments with pretty low demand, thus there could be some extra wind energy available to be exploited.

The graph gives directions on when it is optimal to incentivize consumers to use more appliances or to use less of them, as requested by the commission to the national energy authorities: it suggests pushing the use of devices during the middle of the day while reducing the utilization for the first hours of the morning and late afternoon/early evening. It goes slightly against the current system that incentivizes turning on appliances during the night. The take-home message is thus to change the current policy, which favoured the shift of load at night, when industries were not working, with measures that will increase energy use in the middle of the day.

**2) Which would be Italy's most convenient power system considering different gas prices?**

The trend in natural gas prices has been worrying in the last months, even if they have apparently reached a setback, returning to values around 100€/MWh[1] (which are still very high compared to prices one year ago, seeing that it was 45.9 €/MWh on average in 2021). The reason behind this stop in the costs race is a more precise collective action taken by the European Union and a lower-than-expected demand due to record high temperatures for this autumn. These elevated prices could be a push in the prescribed direction, bearing in mind the perspective of decarbonization in the following years. Of course, today high prices are an issue for families and energy-intensive industries, which will need assistance and subsidies to sustain this critical situation. From a market viewpoint, though, it can be an opportunity to transform the power system into a more cost-effective one. An evaluation of how the system would change according to different gas prices, based on current costs of technologies and without any constraint on CO2 is performed, to give perspective and offer strategic insights. The Oemof model of the Italian power sector can perform iteratively different expansion capacity optimization changing natural gas price. In this case they are varied discontinuously, going from 25 to 300 €/MWh with a step of 25 €/MWh. The high end of these cases are obviously quite extreme, but only slightly exceed the gas price cap recently set by the EU at 275 €/MWh.

- Electricity mix

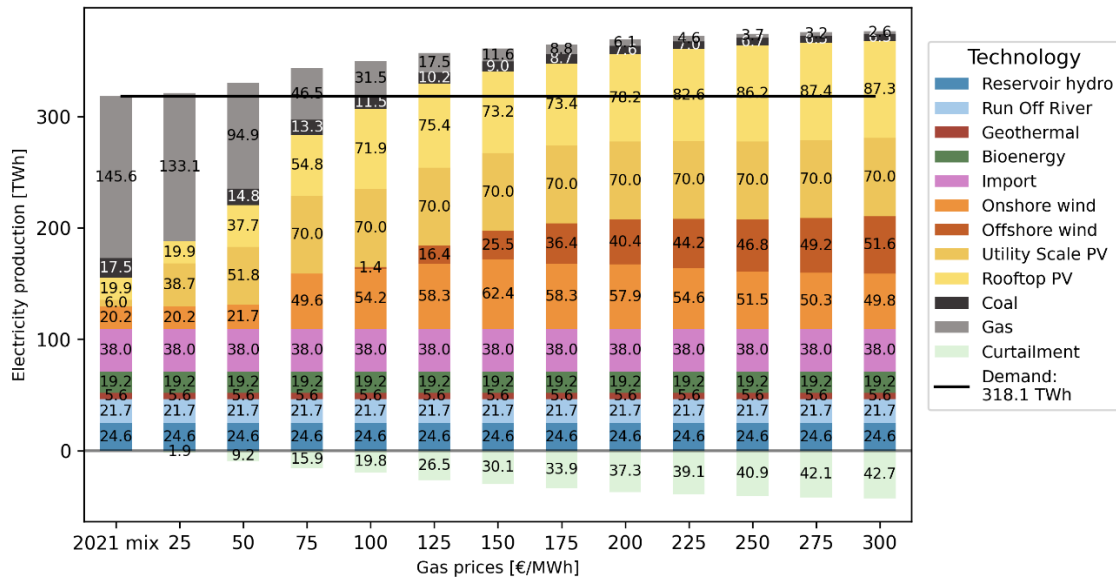

*Figure 5 – Electricity mix for Italy optimized at different gas prices from 25 to 300 €/MWh*

Figure 5 displays the electricity mix optimized for different natural gas price conditions; plus, the current blend is present for comparison in the first column. This fossil resource utilization drops as the prices rise and more renewable sources come into play. Already with a gas price of 50 €/MWh, the cost-minimizing energy system reduces gas consumption by more than one third, which is what the EU projects to be needed to comply with the Fit-for-55 objectives. For low gas price increases solar utility scale energy output is increased, then onshore wind becomes relevant, but offshore turns out to be more cost-efficient for very high prices. In the second column on the left, with 25€/MWh, natural gas is more convenient than coal, assumed at the 2021 average cost of 13.53 €/MWh[36], so it is never employed in the generation. In the negative part of the chart, curtailment is illustrated: excess energy reaches considerable values for higher prices and so for power systems with a high penetration of RES. This is an inevitable downside of a deeply decarbonized electric system, but it can be an excellent opportunity to prompt the decarbonization of other sectors. Curtailed electricity is available practically for free and can be used to electrify end-uses or produce green hydrogen for hard-to-abate industries and for aviation and maritime fuels.

- Installed capacity nationally

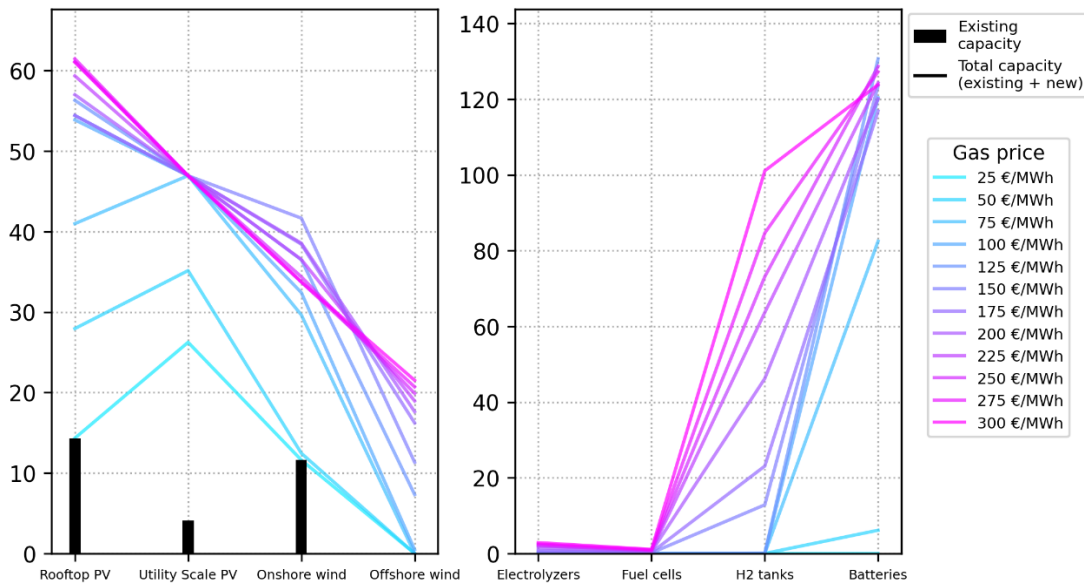

*Figure 6 – Installed capacity at national level per technology*

How much power would be needed for each technology to achieve these optimized power sectors? Figure 6 shows model outputs, highlighting the existing capacities with black bars, while the lines represent the total installed capacity optimized (so current + added) at different prices. Even with the limited cost of 25€/MWh, new power for utility scale photovoltaic is installed, a technology that has already reached market parity and is more convenient than gas. At 50€/MWh, also rooftop PV, onshore wind and batteries are installed and, after 125, hydrogen storage and offshore wind are added. The assumed prices make these two technologies cheaper than using gas with that price, even if they are still quite expensive. Batteries reach a plateau, as seen more clearly in Figure 7, around 120 GWh, while hydrogen storage capacity continuously grows. The latter offers the possibility to the model to independently increase the amount of energy stored and the power to use it, with a large capacity in $H_2$ tanks but a small one for electrolyzers and fuel cells.

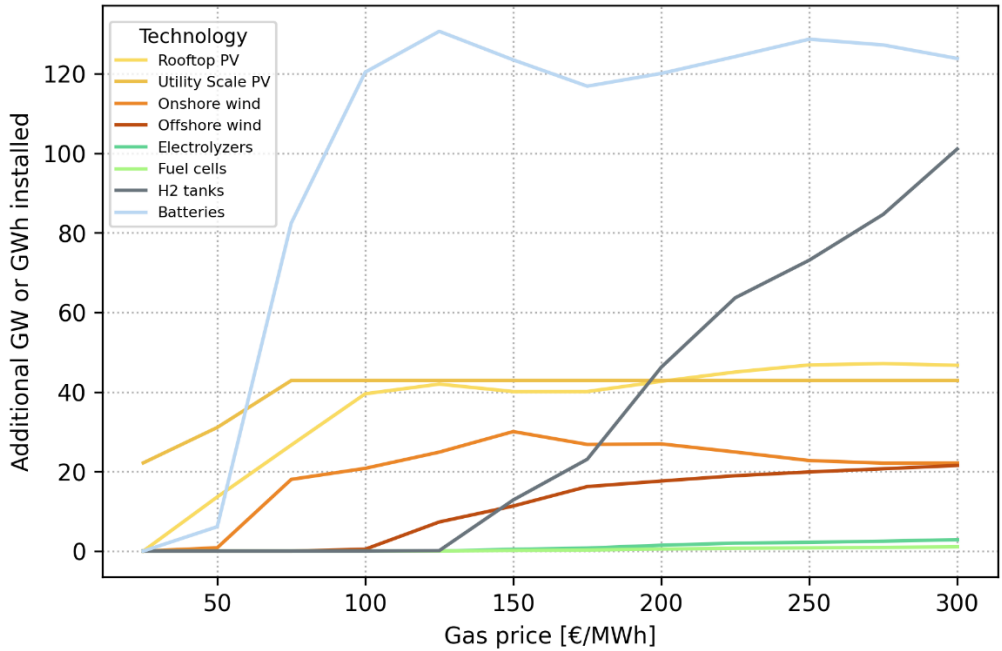

*Figure 7 – Installed capacity at national level per gas price*

- Installed RES capacity per market zone

In Figure 8 the regional detail of the optimized capacity is depicted, where points represent the maximum potential for each technology in a specific region (for rooftop PV it is omitted since it is a substantial value). Besides the two lowest gas costs scenarios, the maximum potential for utility scale photovoltaic power is already met in every market zone, confirming the robust cost-effectiveness of this solution in the energy transition. Offshore wind is a crucial player in the South and Sardinia thanks to its great resource availability in these regions. At the same time, in the North, the main driver for this technology installation is the need to satisfy a very high local demand.

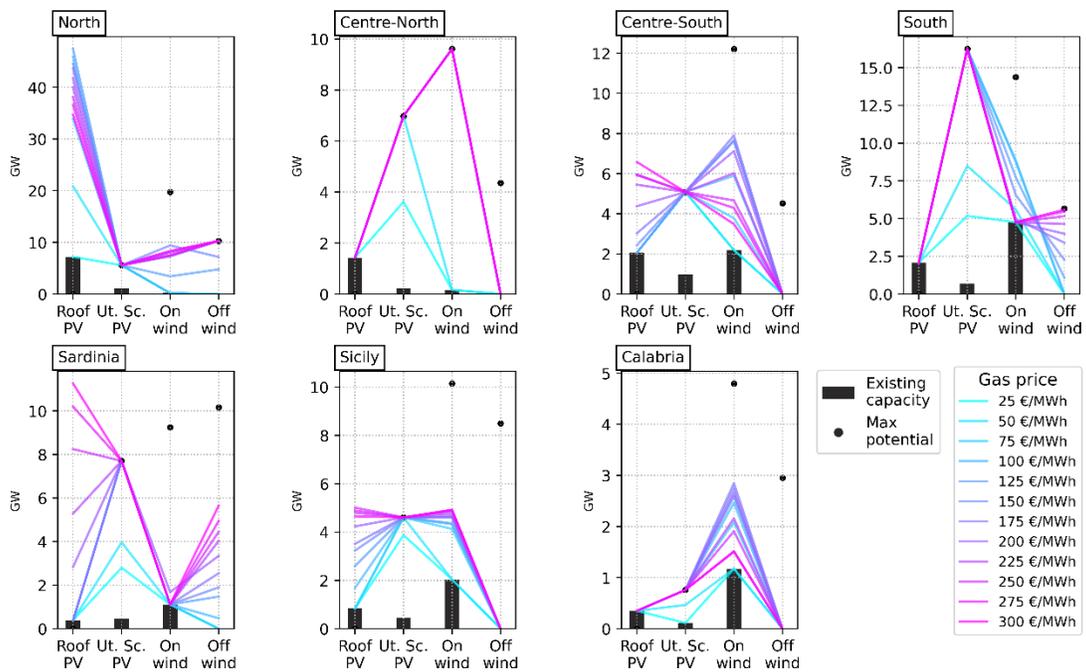

*Figure 8 – Installed capacity at national level per technology. Note the different y-axis scales.*

- Transmission lines, costs and emissions

The Italian power sector model can also expand transmission lines between market zones, providing flexibility, essential in countries like Italy, where renewable resources availability and demand are located distantly in space. The optimizer enhances the most fragile powerlines, as visible in Figure 9. Sardinia to Centre-North is a transmission line that needs significant extra power, with more than 3GW added, to transport a consistent amount of electricity from offshore wind and solar to the peninsula. Terna plans to expand the capacity of this powerline up to 1.1GW[17], starting at 0.4GW with the new project SA.CO.I3, but this graph suggests that reaching deep decarbonization might require more interventions. Centre-North to North capacity is increased already at 50€/MWh; this is also a very important powerline for the stability of the Italian power system: Terna plans to upgrade it from 2.7GW to 3.1GW, while the model optimizes it at values higher than 5GW for 300€/MWh of natural gas. Two other transmission lines require minor adjustments, Sicily to Sardinia and Sicily to Centre-South, values in line with Terna's strategy for constructing the Tyrrenian link[37].

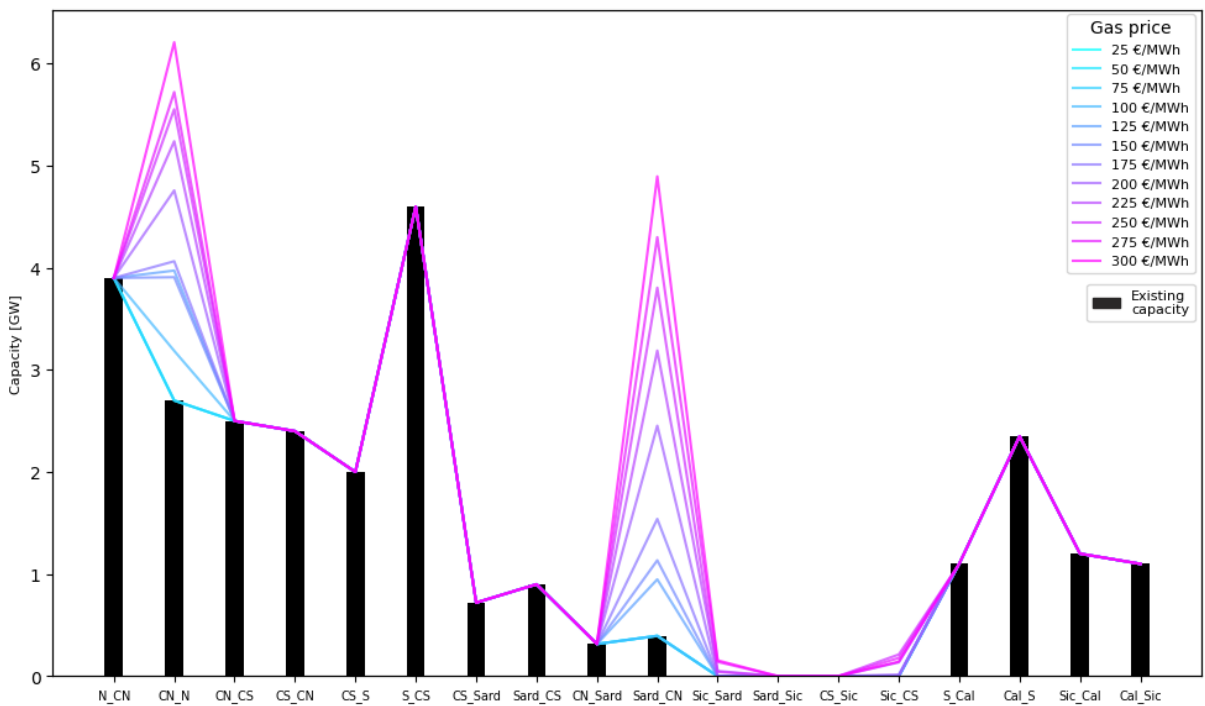

*Figure 9 – Installed capacity for transmission lines between market zones*

For higher gas costs, the total annual costs for the system increase as well, but the trend appears to be logarithmic (see Figure 10). In the model, this parameter represents the total annualized cost to install new capacity plus the operational costs. Installing new RES and storage technologies becomes increasingly more convenient than generating electricity with expensive gas. Emissions show a quite descending trend after 50€/MWh: the first increasing step happens because, with 25€/MWh for gas, no coal is used to produce electricity, so the system has less $CO_2$ emissions. Even for medium prices in these optimized scenarios, carbon dioxide emissions drop to values significantly lower than today (81.1 Mton of CO2 in 2019[34]). High fossil fuel prices make transitioning to a cleaner power system extremely feasible and convenient from an economic point of view. For 300 €/MWh, the model reaches a reduction of 94% of the emissions compared to 1990 levels without imposing any constraint on CO2.

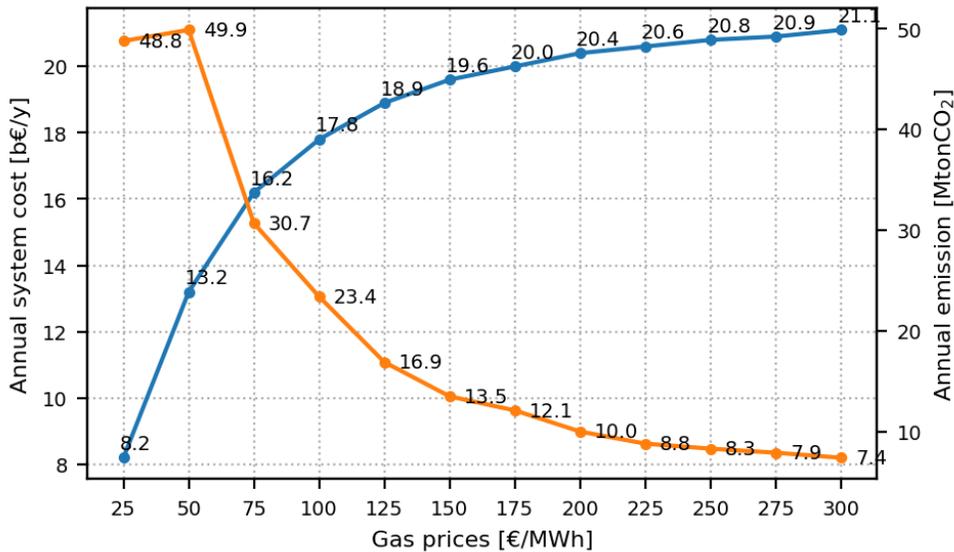

*Figure 10 – Total annual system cost and total yearly emissions of $CO_2$ per gas prices*

### 3) What could be the role of Demand Side Measures in power systems with a high penetration of RES?

Many policies and measures to push the transition of power sectors are applied to the supply side to increase the capacity of renewable energy sources installed or to promote the development of different types of storage. An element often underestimated is the demand, which has a great potential for emission reduction and could play a decisive role in reducing the costs of the decarbonization process[38]. Demand Side Management can shift peak demand from moments with low production from RES to hours when there is a surplus of renewable generation, thus reducing prices, curtailing less energy and using fewer fossil fuels. DSM is a way to provide flexibility, currently with limited impact since the only form of storage widely present is PHS. Instead, it could be a valuable asset in a power system with a high share of RES and storage technologies, decreasing the amount of RES capacity installed for an adequate electricity supply.

To evaluate the pros and cons of demand side measures in deeply decarbonized electric systems, the case study will be the Italian power sector in the year 2030, optimized with the oemof model. The preliminary assumptions are:

- Coal phase-out in 2025[39], so coal-fired power plants are no more available in 2030
- Powerlines capacity assumed according to Terna Development Plan 2021[17]
- Demand increased at 2030 projections following Terna Adequacy Report Terna 2021[40], which expects an annual load of 331 TWh, a 4% growth of 4% with respect to the 2021 value of 318.1 TWh
- Implementing the legally binding objective of the package Fit-for-55[41], which sets a target for reducing total emissions by 55% concerning 1990 levels. For the power sector, a more considerable effort is required, which will be to approximately drop 65% of its emissions[42] (126.4 MtonCO2 in 1990[34], so 44.24 MtonCO2 in 2030)
- Costs of technologies as in 2021 (see Table 2), a conservative assumption since prices tend to go down, but that reflects the idea that investments should start as soon as possible for a near-term transition.

A relevant unknown for these scenarios is the gas of natural gas in 2030, so we will evaluate two hypotheses: 25 and 50 €/MWh. The first one is coherent with the forecast in the World Energy Outlook 2022[38] from IEA, in the Announced Pledges (27 €/MWh) and the Stated Policy cases (29 €/MWh). The second value is in line with the 2021 average price of 45.9 €/MWh[43], representing a mild price shock.

- Electricity mix

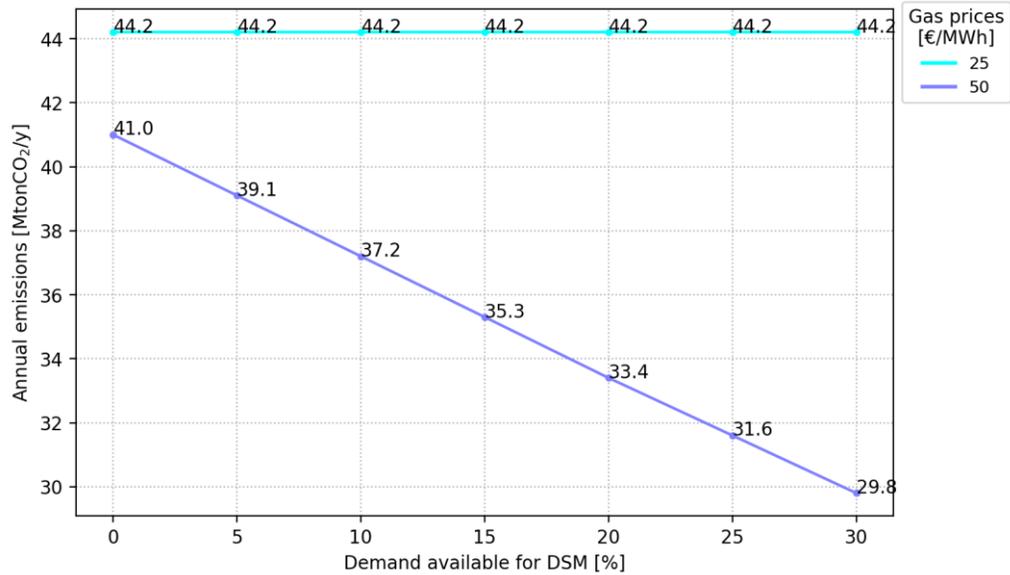

*Figure 11 – Annual $CO_2$ emissions for different % of demand available for DSM*

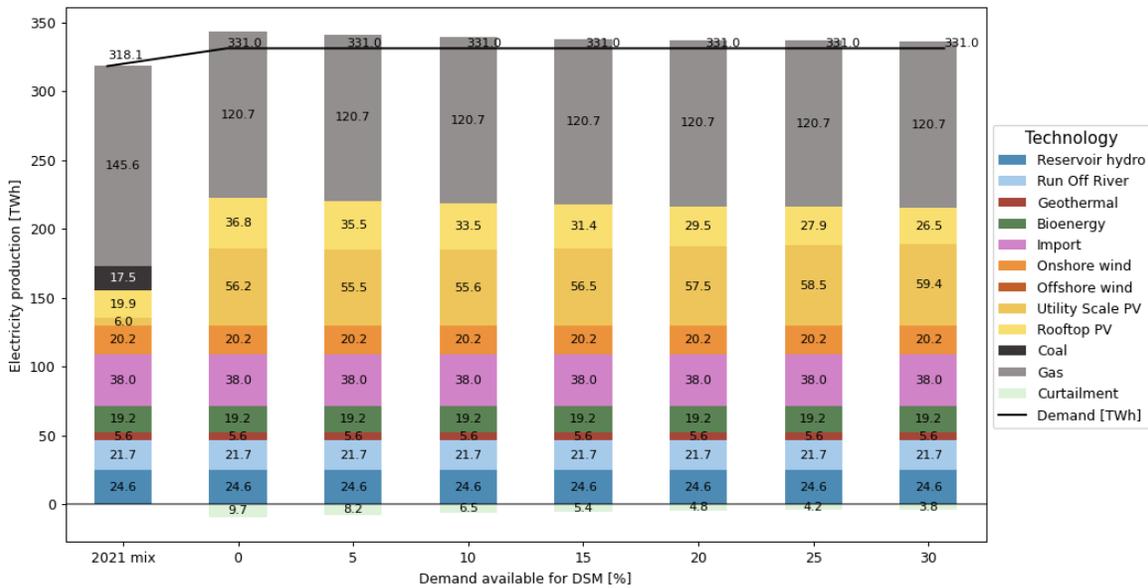

*Figure 12 – Electricity mix for different % demand available for DSM, gas price 25 €/MWh*

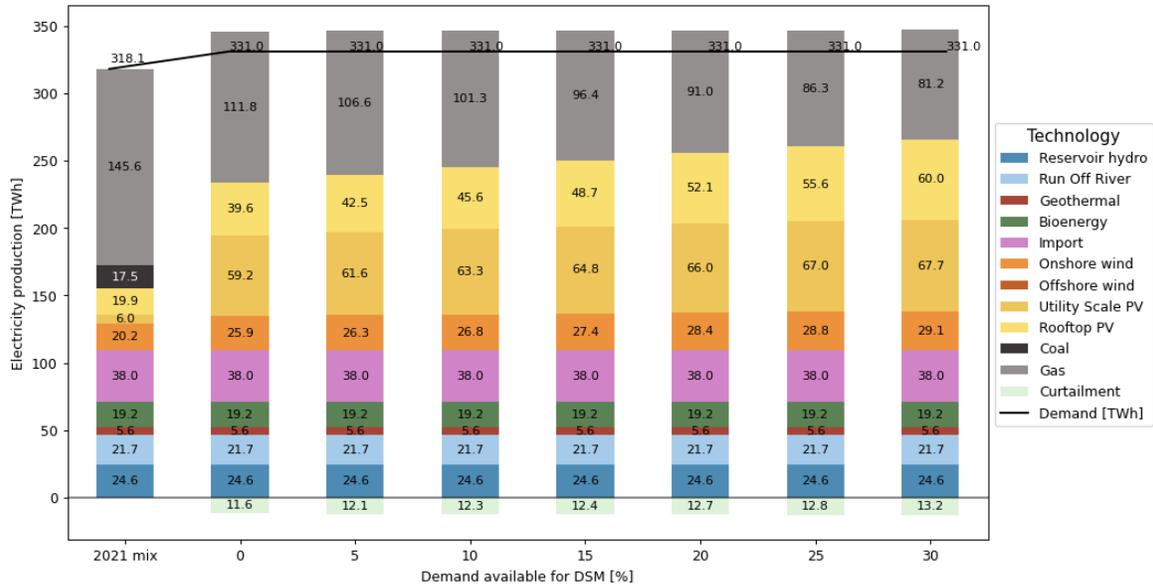

*Figure 13 – Electricity mix for different % demand available for DSM, gas price 50 €/MWh*

The first point to highlight in the results is that, with gas prices at 50 €/MWh, the emission reduction is pushed more than the -65% objective (see Figure 11) since expensive resource prices lead to deeper decarbonization for economic convenience, not for emissions constraint. A power system with -65% emissions would look like the ones in Figure 12 and natural gas is still part of the mix, at around 90% of today's consumption (140TWh of electricity[23]). This suggests that a practical and notable step in the near-term transition of the electricity sector would be to phase-out coal, as also highlighted in a recent IEA report[44], keeping natural gas as a temporary flexibility measure. With high natural gas prices (Figure 13), since electricity is primarily produced by RES capacity, curtailment is a great share of electricity production in all the scenarios. For the 25 €/MWh scenarios, DSM can help reduce up to 60% of the curtailed electricity since it can optimize the demand and push it in a moment of extra renewable generation. For 50 €/MWh instead, the driver for the demand shift is to reduce the system costs by decreasing the use of more expensive natural gas, even if this leads to the production of excess electricity basically for free from the extra RES capacity installed. Eventually, the nationally installed RES capacity across the different cases results always in between 60 and 70% of the total installed generation capacity (see Table 1), thus passing clearly the REPowerEU target of 40-45%[45].

- Installed capacity nationally

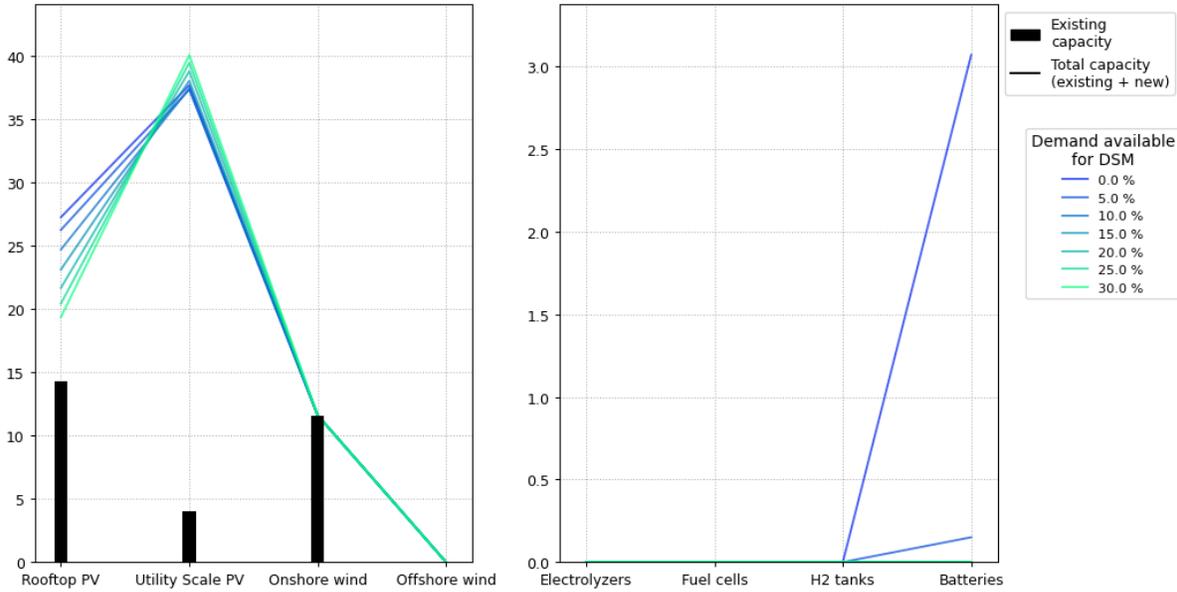

*Figure 14 – Installed capacity nationally for different % demand available for DSM, gas price 25 €/MWh*

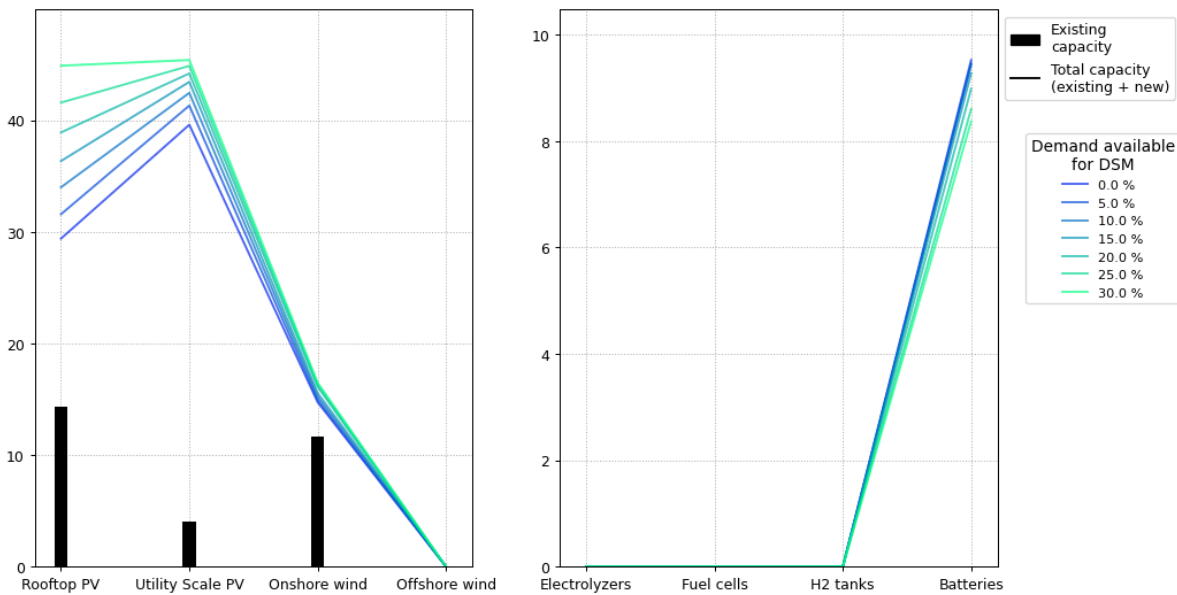

*Figure 15 – Installed capacity nationally for different % demand available for DSM, gas price 50 €/MWh*

In both cases, it is evident that exploiting demand-side measures can lead to a significant reduction in the storage capacity installed (Figure 14 and Figure 15), choosing a larger capacity for renewable generation instead. This helps reduce the system's total costs since, nowadays, solar and wind technologies have a price lower than the ones for batteries or hydrogen storage, which are less mature and cumulatively less installed. As already said, utility scale photovoltaics is currently the most convenient technology. A 65% emission reduction would be reached by installing around 4 GW per year of utility scale, plus around 1 GW per year of rooftop PV. In the first graph, a higher percentage of demand available for DSM cuts down rooftop PV capacity (with equivalent hours slightly lower than utility scale one) and the overall RES newly installed power, even if utility scale moderately increases. Finally

- Installed RES capacity per market zone

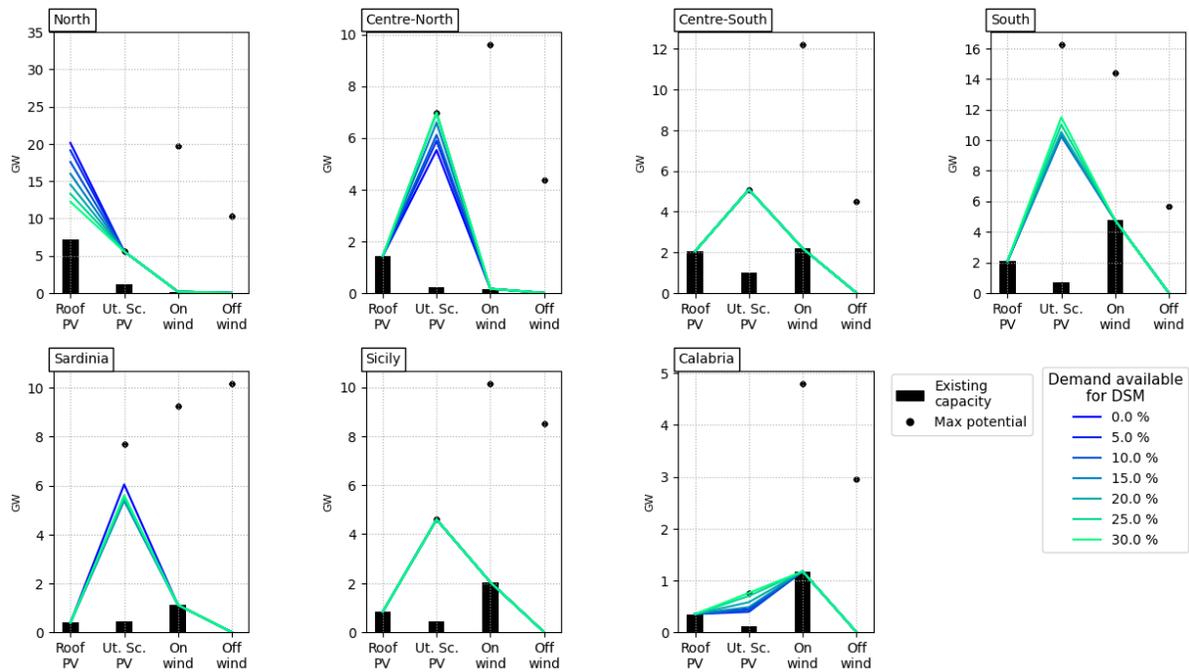

*Figure 16 - Renewable capacity installed per market region for different % demand available for DSM, gas price 25 €/MWh*

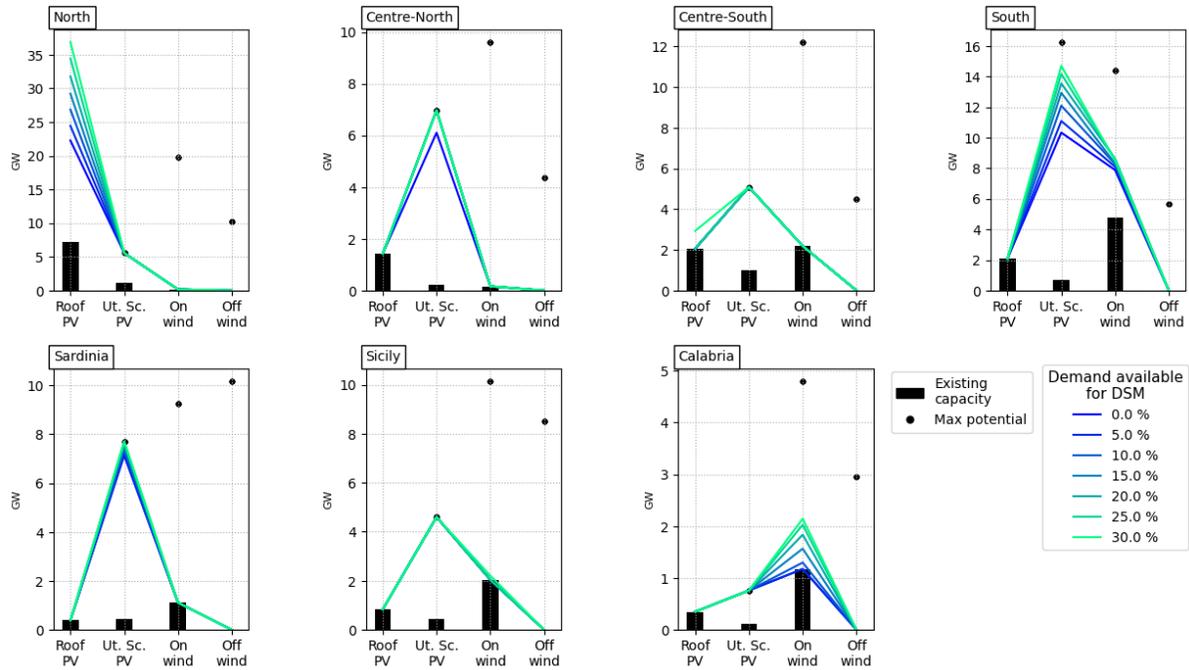

*Figure 17 - Renewable capacity installed per market region for different % demand available for DSM, gas price 50 €/MWh*

Both in Figure 16 and Figure 17, utility scale PV capacity reaches its max potential in almost every scenario and region, except in the South where its potential has quite a considerable absolute value. For low gas costs, the new power for utility scale solar decreases with a larger percentage of DSM allowed since it works towards optimizing

RES utilization and reducing curtailment. For 50 €/MWh, a higher share of demand shift brings more rooftop PV in the North (where there is a big demand) and more utility scale PV in Centre-North, South, Calabria (where sun availability is greater), confirming that in this case DSM has the role of decreasing gas utilization.

In these scenarios, transmission lines have the power prospected in the Development Plan 2021 from Terna and no additional capacity is installed; thus, the planned expansions could be enough to provide the correct connection and flexibility to the power system.

- Costs

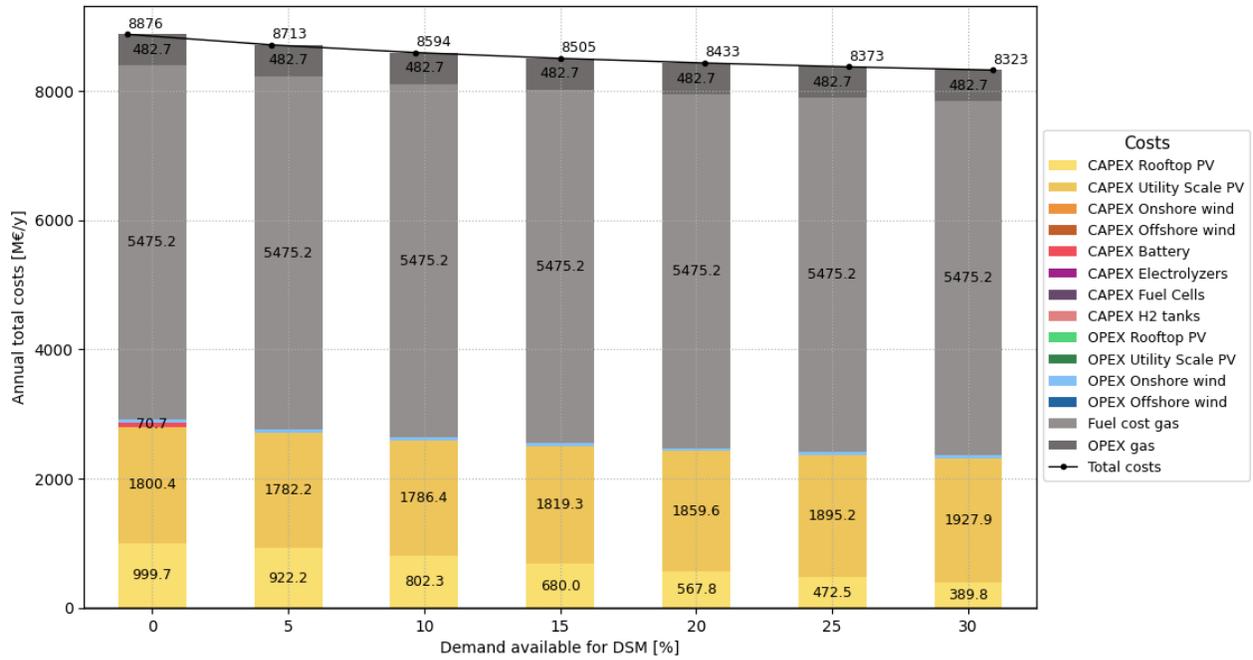

*Figure 18 – Cost items for different % demand available for DSM, gas price 25 €/MWh*

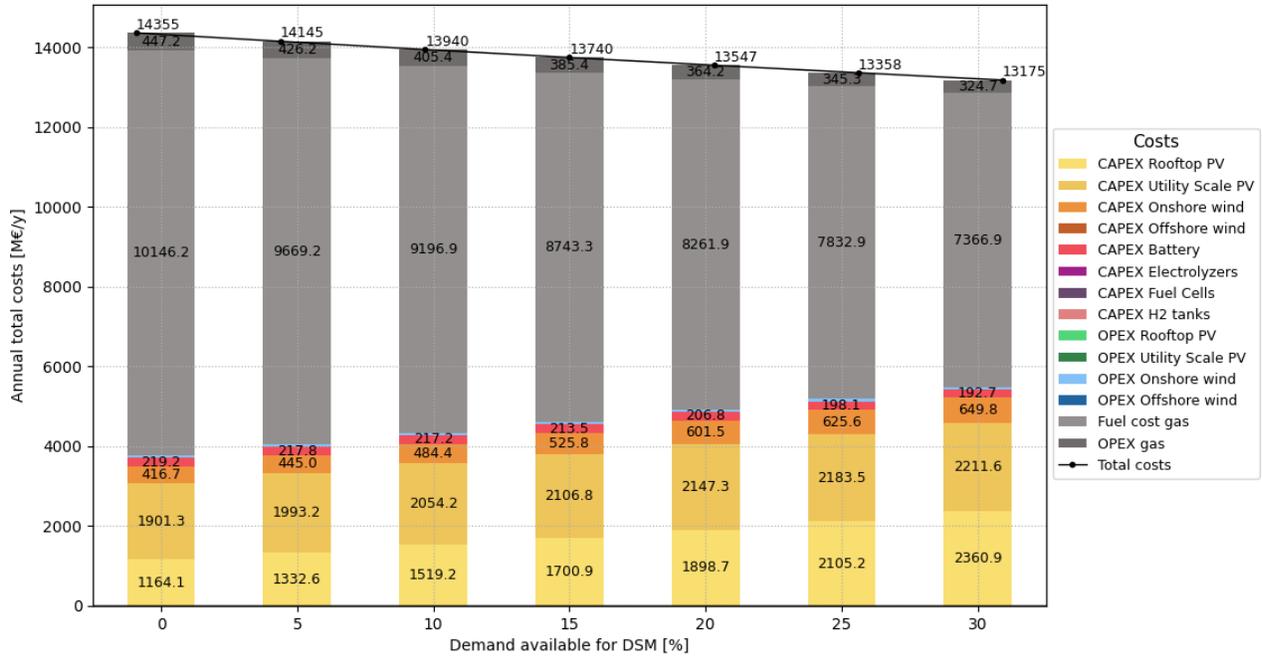

*Figure 19 - Cost items for different % demand available for DSM, gas price 50 €/MWh*

Annual system costs go down as the DSM percentage goes up, in the first case (Figure 18) thanks to a lower amount of installed rooftop PV and in the second one (Figure 19) mainly due to a reduction in costs for the purchase of natural gas (up to almost -30%). For high gas prices, the cost item for the capital invested in lithium-ion batteries shrinks for larger shares of demand shift allowed.

- Hourly demand

To show the effect of DSM and how the hourly generation is forecasted to be in 2030, two typical days are displayed for Sardinia with 0% DSM and 30% DSM, in the 50 €/MWh case (Figure 20). This region is selected because it has the largest battery capacity and will have a massive extra RES power that will change the electricity mix.

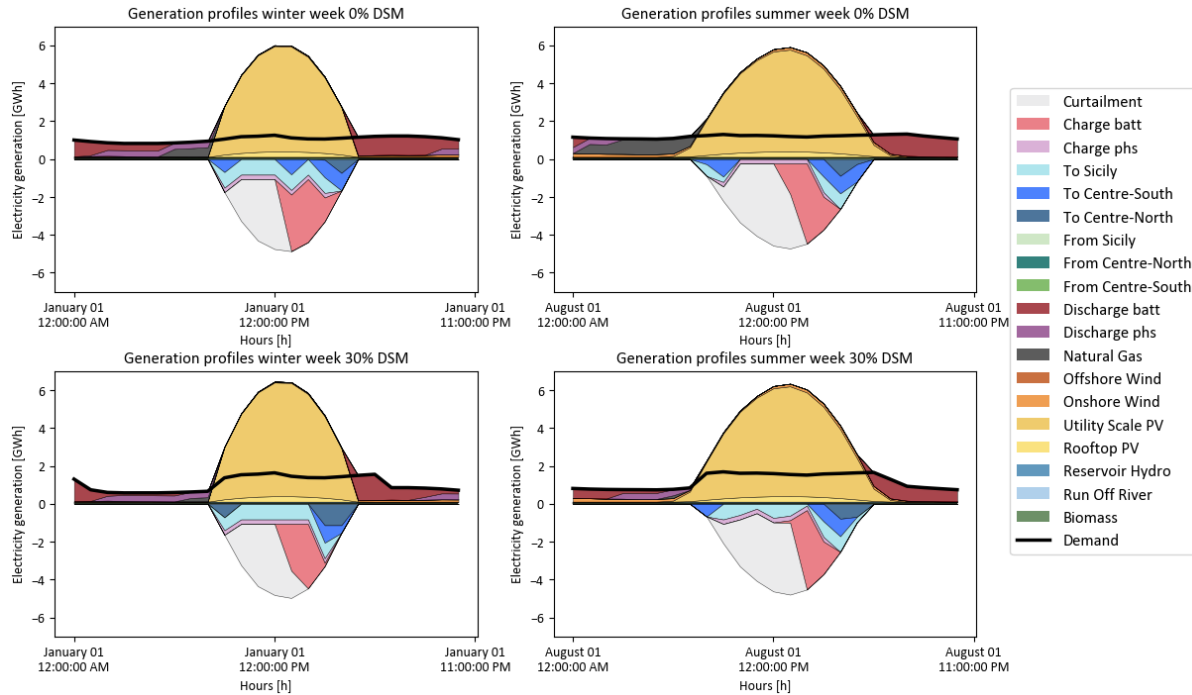

*Figure 20 – Hourly generation for typical winter (left) and summer (right) days in Sardinia, cases with 0% demand available for DSM (up) and 30% (down), gas price 50 €/MWh*

It is interesting to notice that the demand with 30% DSM is lower for night hours and higher for day hours, greatly decreasing the amount of gas required. Exploiting DSM, storage utilization is lower and curtailment is larger, even if excess electricity is a significant quantity also in the 0% cases, since the system has a high amount of solar electricity production. In all the scenarios, the role of Sardinia as an electricity producer and exporter to the rest of the peninsula is noticeable, thanks to its relevant sun and wind availability.

## 4. Conclusions and policy-relevant considerations

In this paper, a power system model for Italy is exploited to evaluate the impact of European policies, the dynamics generated by higher gas prices and the usefulness of demand side measures in the 2030 electricity sector. The first analysis highlights the implementation of the European Council directive for the months between December 2022 and March 2023, trying to alleviate the burden of natural gas prices on families and firms and to reduce the import of this resource from Russia. This measure consists of two steps, starting with a 5% load shift in 10% of the electric demand peak hours. This has the effect of decreasing the utilization of the only source of flexibility in the current system, which is PHS (-85% utilization), and probably reduces the urgency of the demand for gas in electricity, thus stabilizing the volatility of prices. The second step is a voluntary 10% demand reduction for these four months, which leads to a decline in gas consumption of 20% and so in emissions, thanks to a boost in PHS energy output. With these two measures together, it is possible to identify the best hours to shift the load up or down: the first ones are in the middle of the day, when there is a large availability of sun, while the last ones are in the early morning and late afternoon. The advice for decision makers is to push more for demand reduction measures and to offer incentives for electricity consumers to use appliances or run industrial processes between around 11 and 14.

The second study explores the optimal Italian power system resulting from investment optimizations at different gas prices, going from 25 to 300 €/MWh. Outcomes point out strongly that the high costs of natural gas do not go against energy transition, they favor it more since renewable technologies will be way cheaper than producing electricity with gas power plants. These results highlight a pivotal matter for policymakers: creating policies to cap artificially natural gas prices can lead to a distortion in the market, favoring a major use of this resource; even if you plan them as a temporary measure, it can be complex to remove them from a social acceptance point of view, similar to what happened in Brazil with energy subsidies. This behavior goes against the recent climate agreement with the European Union (Fit-for-55), requiring a reduction of emissions of 55% in the year 2030[41]. A solution to alleviate the economic burden on the population could be to provide a temporary bonus or lump-sum incentives, that are technology neutral and can help families and businesses to stand sudden price shocks. For long-term resilience, it is suggested to set up frameworks for targeted incentives that can reduce energy consumption, particularly gas consumption. Governments could offer economic stimuli for energy efficiency, for building heat pumps, converting industry processes, and purchasing electric cars.

The third part of the work investigates how demand side management could offer opportunities for cost reduction in power systems with a high penetration of RES, in this case in Italy in 2030 (with coal phase-out and a constraint of carbon dioxide emissions at -65% of 1990 values). The first thrilling output is that to reach the emission goal for 2030, the most effective maneuver would be to eliminate coal power plants, followed by the installation of economically convenient utility scale PV, while natural gas could still represent around 35% of the electricity mix. Regarding DSM, its role could be twofold: with low gas prices, it is crucial to reduce the curtailed electricity since it can tune the load optimally, shifting it when there is excess RES generation; with high gas prices, its effect would be to avoid the utilization of this expensive resource when it is possible, thus decreasing the overall costs of the system. In both cases, DSM helps diminish the capacity installed of storage technologies, which are still not extremely mature. Demand side measures are challenging to implement since they require strategic social policies to make them accepted and convenient for citizens. Still, it is worth exploring these solutions to reduce costs and storage capacity needed.

**List of acronims**

| | |
|---|---|
| CN | Centre-North |
| CS | Centre-South |
| DIW | Deutsches Institut für Wirtschaftschung (German Institute for Economic Research) |
| DLR | Deutschen Zentrum für Luft- und Raumfahrt (German Aerospace Center) |
| DSM | Demand Side Management |
| EC | European Council |
| EU | European Union |
| IEA | International Energy Agency |
| N | North |
| PHS | Pumped Hydro Storage |
| PV | Photovoltaics |
| RES | Renewable Energy Sources |
| S | South |

**References**

1. EU Natural Gas - 2022 Data - 2010-2021 Historical - 2023 Forecast - Price - Quote. https://tradingeconomics.com/commodity/eu-natural-gas.

2. External Sector Statistics | Bank of Russia. https://www.cbr.ru/eng/statistics/macro_itm/svs/.

3. Quarterly report on European gas markets_Q4 2021.pdf.

4. Importazioni di gas naturale - Analisi e statistiche energetiche e minerarie - Ministero della Transizione Ecologica. https://dgsaie.mise.gov.it/importazioni-gas-naturale.

5. Where does the EU's gas come from? https://www.consilium.europa.eu/en/infographics/eu-gas-supply/.

6. Council agrees on emergency measures to reduce energy prices. https://www.consilium.europa.eu/en/press/press-releases/2022/09/30/council-agrees-on-emergency-measures-to-reduce-energy-prices/.

7. News, B. Europe's Energy Crisis Complicates COP for Key Climate Player - BNN Bloomberg. *BNN* https://www.bnnbloomberg.ca/europe-s-energy-crisis-complicates-cop-for-key-climate-player-1.1846063 (2022).

8. Hilpert, S. *et al.* The Open Energy Modelling Framework (oemof) - A new approach to facilitate open science in energy system modelling. *Energy Strategy Rev.* **22**, 16–25 (2018).

9. Le nuove zone del mercato elettrico: quello che c'è da sapere | Lightbox. *Terna Energy Blog* https://lightbox.terna.it/en/insight/new-electricity-market-zones.

10. ALLEGATO A.24 AL CODICE DI RETE: INDIVIDUAZIONE ZONE DELLA RETE RILEVANTE - PDF Free Download. https://docplayer.it/9456044-Allegato-a-24-al-codice-di-rete-individuazione-zone-della-rete-rilevante.html.

11. Di Bella, A. *et al.* Multi-Objective Optimization to Identify Carbon Neutrality Scenarios for the Italian Electric System. SSRN Scholarly Paper at https://doi.org/10.2139/ssrn.4134221 (2022).

12. oemof.solph package — oemof.solph 0.4.4 documentation. https://oemof-solph.readthedocs.io/en/stable/reference/oemof.solph.html#module-oemof.solph.custom.sink_dsm.

13. Gils, H. C. Balancing of intermittent renewable power generation by demand response and thermal energy storage. (2015) doi:10.18419/OPUS-6888.

14. Zerrahn, A. & Schill, W.-P. On the representation of demand-side management in power system models. *Energy* **84**, 840–845 (2015).

15. Prina, M. G. *et al.* Multi-objective investment optimization for energy system models in high temporal and spatial resolution. *Appl. Energy* **264**, 114728 (2020).

16. Download center - Terna spa. https://www.terna.it/en/electric-system/transparency-report/download-center.

17. Piano di Sviluppo 2021. 372.

18. Mongird, K. *et al.* 2020 Grid Energy Storage Technology Cost and Performance Assessment. 117 (2020).

19. Energy System Technology Data. (2022).

20. Batas Bjelić, I. & Rajaković, N. Simulation-based optimization of sustainable national energy systems. *Energy* **91**, 1087–1098 (2015).

21. Tröndle, T., Pfenninger, S. & Lilliestam, J. Home-made or imported: On the possibility for renewable electricity autarky on all scales in Europe. *Energy Strategy Rev.* **26**, 100388 (2019).

22. Le Centrali in Italia – Assocarboni. https://assocarboni.it/assocarboni/il-carbone/le-centrali-in-italia/.

23. Pubblicazioni Statistiche - Terna spa. https://www.terna.it/it/sistema-elettrico/statistiche/pubblicazioni-statistiche.

24. Fonti rinnovabili - Terna spa. https://www.terna.it/it/sistema-elettrico/dispacciamento/fonti-rinnovabili.

25. JRC Hydro-power plants database. (2022).

26. Costs of utility-scale photovoltaic systems integration in the future Italian energy scenarios - Veronese - 2021 - Progress in Photovoltaics: Research and Applications - Wiley Online Library. https://onlinelibrary.wiley.com/doi/full/10.1002/pip.3382.

27. Rodríguez-Martinez, Á. & Rodríguez-Monroy, C. Economic Analysis and Modelling of Rooftop Photovoltaic Systems in Spain for Industrial Self-Consumption. *Energies* **14**, 7307 (2021).

28. Renewable Power Generation Costs in 2021. https://irena.org/publications/2022/Jul/Renewable-Power-Generation-Costs-in-2021.

29. Average pack price of lithium-ion batteries and share of cathode material cost, 2011-2021 – Charts – Data & Statistics. *IEA* https://www.iea.org/data-and-statistics/charts/average-pack-price-of-lithium-ion-batteries-and-share-of-cathode-material-cost-2011-2021.

30. 2021 CMTS. *Tableau Software* https://public.tableau.com/views/2021CMTS/TechSummary?:embed=y&Technology=Utility-Scale%20Battery%20Storage&:embed=y&:showVizHome=n&:bootstrapWhenNotified=y&:apiID=handler0.

31. Christensen, A. Assessment of Hydrogen Production Costs from Electrolysis: United States and Europe. 73.

32. Cigolotti, V., Genovese, M. & Fragiacomo, P. Comprehensive Review on Fuel Cell Technology for Stationary Applications as Sustainable and Efficient Poly-Generation Energy Systems. *Energies* **14**, 4963 (2021).

33. James, B. D., Houchins, C., Huya-Kouadio, J. M. & DeSantis, D. A. *Final Report: Hydrogen Storage System Cost Analysis*. https://www.osti.gov/biblio/1343975 (2016) doi:10.2172/1343975.

34. r343-2021.pdf.

35. 5-PRODUZIONE_8d9cecf70a531dd.pdf.

36. International Monetary Fund. Global price of Coal, Australia. *FRED, Federal Reserve Bank of St. Louis* https://fred.stlouisfed.org/series/PCOALAUUSDM (1980).

37. Tyrrhenian link: il doppio collegamento sottomarino tra Sicilia, Sardegna e penisola - Terna spa. https://www.terna.it/it/progetti-territorio/progetti-incontri-territorio/Tyrrhenian-link.

38. Birol, D. F. World Energy Outlook 2022. 524.

39. PNIEC_finale_17012020.pdf.

40. Rapporto Adeguatezza Italia 2021. 52.

41. Fit for 55. https://www.consilium.europa.eu/en/policies/green-deal/fit-for-55-the-eu-plan-for-a-green-transition/.

42. content/enel-com/en/authors/simone-mori. 'Fit for 55': EU on track for climate goals and sustainable growth. https://www.enel.com/company/stories/articles/2021/08/fit-for-55-europe-energy-transition-goals.

43. Global price of Natural gas, EU (PNGASEUUSDM) | FRED | St. Louis Fed. https://fred.stlouisfed.org/series/PNGASEUUSDM.

44. Coal in Net Zero Transitions: Strategies for rapid, secure and people-centred change. 224.

45. REPowerEU: affordable, secure and sustainable energy for Europe. *European Commission - European Commission* https://ec.europa.eu/info/strategy/priorities-2019-2024/european-green-deal/repowereu-affordable-secure-and-sustainable-energy-europe_en.